\documentclass[pdflatex,sn-mathphys-num]{sn-jnl}

\usepackage{graphicx}%
\usepackage{multirow}%
\usepackage{amsmath,amssymb,amsfonts}%
\usepackage{amsthm}%
\usepackage{mathrsfs}%
\usepackage[title]{appendix}%
\usepackage{xcolor}%
\usepackage{textcomp}%
\usepackage{manyfoot}%
\usepackage{booktabs}%
\usepackage{algorithm}%
\usepackage{algorithmicx}%
\usepackage{algpseudocode}%
\usepackage{listings}%

\theoremstyle{thmstyleone}%
\theoremstyle{thmstyletwo}%

\theoremstyle{thmstylethree}%

\begin{document}

\title[Thermoelectric effect at the quantum Hall-superconductor interface]{Thermoelectric effect at the quantum Hall-superconductor interface}


\author*[1]{\fnm{Jordan T.} \sur{McCourt}}\email{jordan.mccourt@duke.edu}

\author[1]{\fnm{John} \sur{Chiles}}

\author[1]{\fnm{Chun-Chia} \sur{Chen}}

\author[2]{\fnm{Kenji} \sur{Watanabe}}

\author[2]{\fnm{Takashi} \sur{Tanaguchi}}

\author[3]{\fnm{Fran\c{c}ois} \sur{Amet}}

\author*[1]{\fnm{Gleb} \sur{Finkelstein}}\email{gleb@duke.edu}

\affil[1]{\orgdiv{Department of Physics}, \orgname{Duke University}, \orgaddress{\street{120 Science Drive}, \city{Durham}, \postcode{27710}, \state{NC}, \country{USA}}}

\affil[2]{\orgdiv{Advanced Materials Laboratory}, \orgname{National Institute for Materials Science}, \orgaddress{\city{Tsukuba}, \postcode{305-0044}, \country{Japan}}}

\affil[3]{\orgdiv{Department of Physics and Astronomy}, \orgname{Appalachian State University}, \city{Boone}, \postcode{28608}, \state{NC}, \country{USA}}


\abstract{The interfaces of quantum Hall insulators with superconductors have emerged as a promising platform to realise interesting physics that may be relevant for topologically protected quantum computing. However, these interfaces can host other effects which obscure the detection of the desired excitations. Here we present measurements of the thermoelectric effect at the quantum Hall-superconductor interface. We explain the heat transport by considering the formation of a hotspot at the interface, which results in a non-equilibrium distribution of electrons that can propagate across the superconductor through vortex cores. The observed thermoelectric effect results in a voltage which changes sign on quantum Hall plateaus and responds to the rearrangement of vortices in the wire. These observations highlight the complex interplay of thermal and charge phenomena at the quantum Hall -- superconductor interfaces and should be considered when interpreting transport measurements in similar systems.}




\maketitle

\section{Introduction}

Hybrid devices combining  superconductors and quantum Hall (QH )insulators have attracted much attention over the past decade~\cite{amet_supercurrent_2016,lee_inducing_2017,seredinski_quantum_2019,zhao_interference_2020,hatefipour_induced_2022,gul_andreev_2022,vignaud_evidence_2023,barrier_one-dimensional_2024,zhao_nonlocal_2024,akiho_andreev_2024,hatefipour_andreev_2024,jang_engineering_2024,villani_quasi-0-periodic_2025}. This effort has been driven by the search for exotic interfacial states, but real samples often demonstrate more complex and less-than-ideal behaviors~\cite{manesco_mechanisms_2022,kurilovich_disorder-enabled_2023,kurilovich_criticality_2023}. One source of this non-ideality is the presence of the vortices in the superconductor, which can absorb the edge states~\cite{zhao_loss_2023}. We have recently demonstrated that hot spots created in the QH samples under bias can heat the electrons inside the core of the vortices, resulting in heat propagating deep into the superconducting contacts~\cite{zhao_thermal_2025}. Here, we explore this phenomenon directly in QH samples bisected by narrow superconducting contacts, in which the heat generated on one side of the contact induces the thermoelectric phenomena on the opposite side.

\section{Results}

\begin{figure}[h]
\centering
\includegraphics[width=\textwidth]{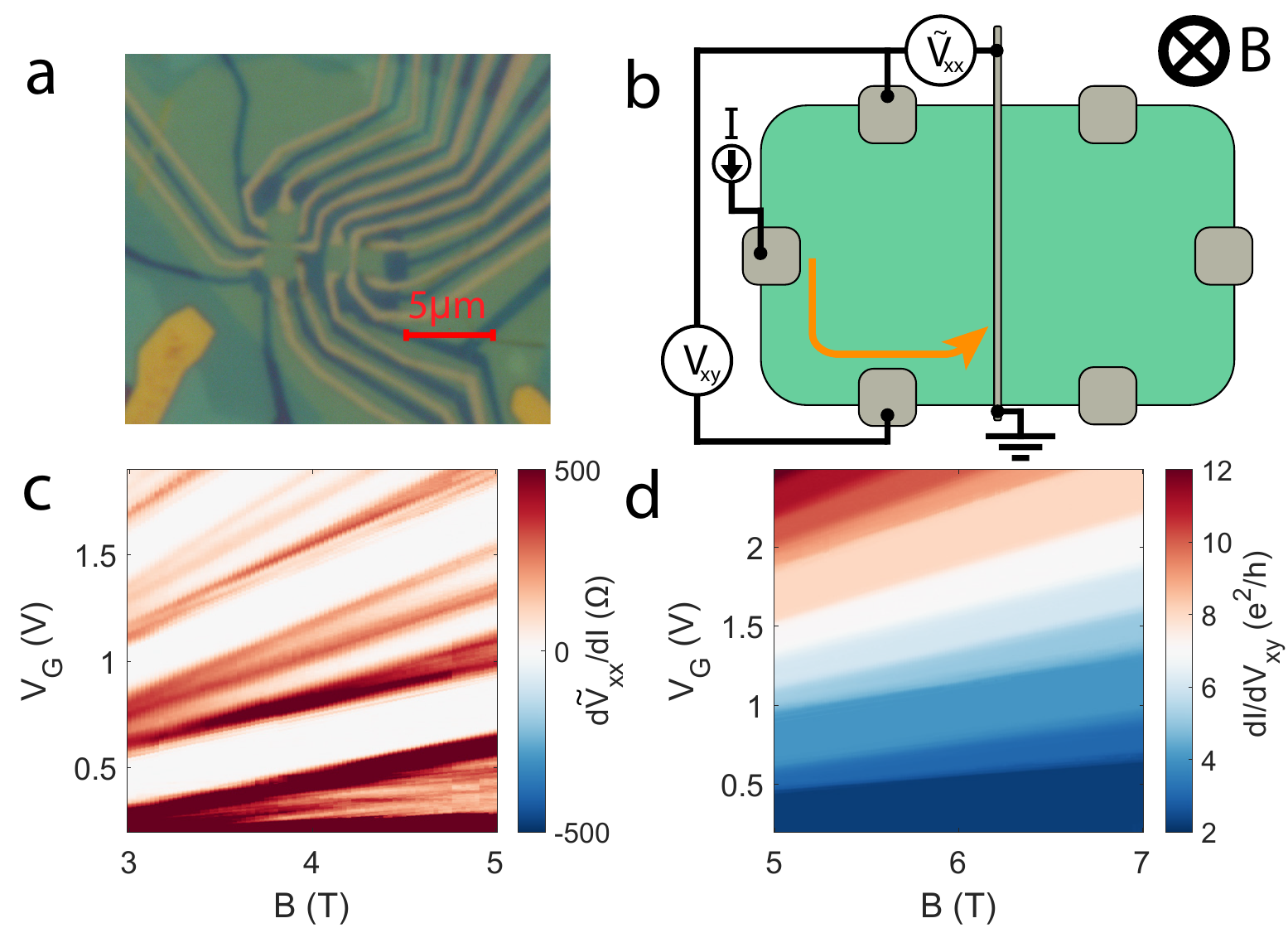}
\label{fig:DeviceFig}
\caption{\textbf{(a)} An optical image of the sample. The 50nm device is on the left and the 250nm is on the right. \textbf{(b)} Device schematic and configuration of the measurements presented in \textbf{c} and \textbf{d}. The orange arrow denotes the direction of travel of electrons when the device is n-doped ($V_G>0$). \textbf{(c)} Measurement of the `downstream' resistance, $\tilde{R}_{xx}$, as a function of magnetic field and gate voltage in the 50nm device. \textbf{(d)} Transverse conductance, $(\frac{dV_{xy}}{dI})^{-1}$, in the 50nm device as a function of magnetic field and gate voltage.}
\end{figure}

\noindent The two devices presented in this work each consist of a Hall bar bisected by a 1~$\mu$m long superconducting contact. These central contacts are shaped as narrow wires of widths 50 nm and 250 nm. An optical image of the device is presented in Figure~1a and a schematic of the measurement configuration is displayed in Figure~1b. The bisecting wire and all other contacts are composed of MoRe, a superconductor with an upper critical field of $H_{c2}\approx10$T and a critical temperature of $T_c\approx10$K. The Hall bars are both patterned in the same piece of bilayer graphene which is encapsulated in hexagonal Boron Nitride (hBN). A global graphite backgate controls the carrier density via gate voltage, $V_G$. The device is measured in the QH regime at a temperature of 60 mK. The high quality of the device is confirmed by the presence of broken symmetry states at relatively low fields (Figure~1d). This ensures that the dominant contribution of transport on quantum Hall plateaus is due to the edge states, rather than through the bulk. \\
\\
Our prior studies of similar QH-superconductor samples focused on the propagation and detection of the Chiral Andreev Edge States (CAES) formed at the interface~\cite{zhao_interference_2020}. Specifically, we measured the non-local voltage induced on the contact downstream of the grounded superconductor, Figure~1b. The corresponding differential `downstream' resistance ($\tilde{R}_{xx}$) demonstrated microscopic fluctuations as a function of the gate voltage. $\tilde{R}_{xx}$ reached comparable positive and negative values depending on whether an incident electron is emitted downstream as a normally-reflected electron or Andreev-reflected hole, respectively. The present devices are designed to suppress this effect by making all superconducting interfaces longer than $\sim 1 \mu$m. As a result, we consistently observe $\tilde{R}_{xx}=0$ (Figure~1c), while the sample remains on the QH plateau (Figure~1d) 
~\cite{zhao_loss_2023}.\\
\\

\begin{figure}[h]
\centering
\includegraphics[width=\textwidth]{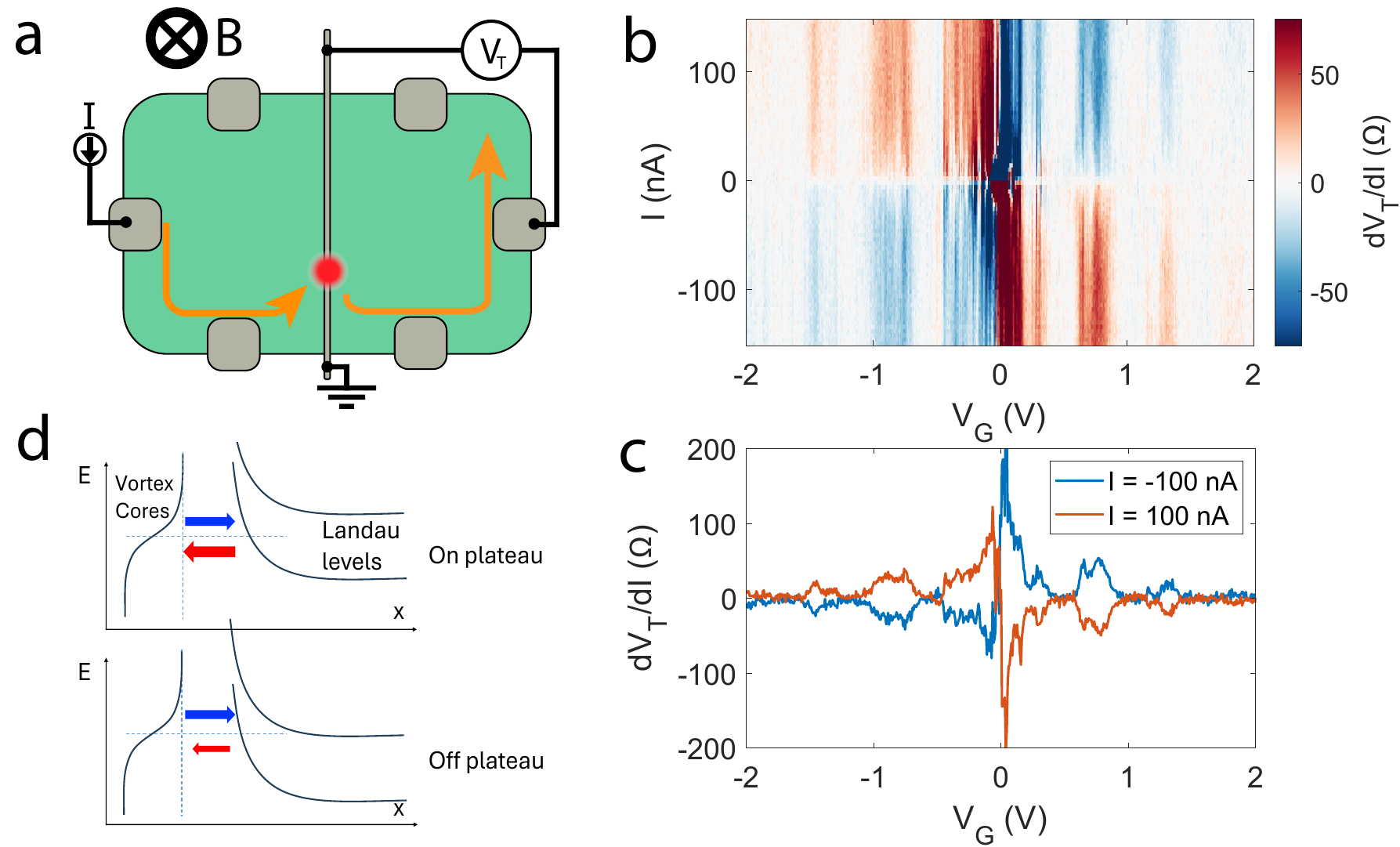}
\label{fig:heating}
\caption{\textbf{(a)} Measurement configuration used in Figures~2-4. Orange arrows denote the direction of electron travel. The hotspot is indicated by the red circle. \textbf{(b)} Map of the differential nonlocal resistance ($R_T=\frac{dV_T}{dI}$) measured vs bias $I$ and $V_G$ in the 50~nm device. The resistance signal is strong between the plateaus and clearly changes sign upon changing the sign of either $I$ and $V_G$. \textbf{(c)} Traces from \textbf{b} at constant $I=100$ nA displaying the dependence of the sign of $\frac{dV_T}{dI}$ on $I$ and $V_G$. \textbf{(d)} Schematic energy diagram showing how the non-equilibrium distribution in the metal exchanges electrons with the Landau levels in graphene. The situation is different on-plateau and off-plateau, in which case the level in the bulk should be preferentially filled with electrons (for n-doping).}
\end{figure}

\noindent We now measure the voltage across the right side of the device. With current bias applied on the left side of the device, the natural expectation would be that all incoming current is shunted by the grounded central contact. Instead, we observe a measurable potential difference between a contact on the right side and the central  contact, $V_T$ (see the Schematic in Figure~2a). The non-local differential resistance  $R_T=\frac{dV_T}{dI}$ is presented in Figures~2b,d. In Figure~2b, we observe that on the QH plateaus $R_T$ is greatly suppressed. At the transitions between the plateau, $R_T$ depends on the sign of both $I$ and $V_G$. For example, $R_T$ is negative when both $I$ and $V_G$ are positive, but it changes the sign when either of them changes sign. This dependence is displayed clearly in Figure~2d.\\
\\
\noindent We explain this observation by the formation of a hotspot at the superconducting interface. The applied current in the 10s of nA range results in a voltage drop on the 100's of $\mu$V scale at the hot spot. We have previously demonstrated that the Joule heating power in the range of 10s of pW can increase the electron temperature at the interface by 100s of mK. That heat can efficiently propagate through a superconducting film at high magnetic field, over distances greatly exceeding the coherence length, presumably due to the overlap of the electrons states in vortex cores~\cite{zhao_thermal_2025,peltonen_magnetic-field-induced_2011}. The non-equilibrium distribution of charge carriers then propagates to the unbiased side of the device, where it can be detected by voltage probes. When $I>0$ and $V_G>0$, `hot' electrons fill the unoccupied states in the bulk as $V_G$ is tuned between QH plateaus. The resulting voltage depends on the applied power, but not the sign of the bias current, resulting in  $V_T(I)$ curves which are peculiarly even in bias. As a result, the non-local differential resistance curves $R_T(I)$ are \emph{odd} in $I$, as seen in Figure~2b,d. For negative $V_G$, the electrons are replaced by holes, which results in the thermal voltage changing sign, as seen in the data.\\
\\
Figure~2 shows that the $R_T$ signal is greatly suppressed on the plateaus. The explanation is likely that the QH edge states have a constant density of states and the hot metal populates them evenly with electrons and holes, see schematics in Figure~2c. However, the signal on the plateau is not exactly zero,  as can be seen in Figure~3a. The individual $V_T$ curves retain the even dependence on $I$, confirming their thermal origin (Figure~3b).
In contrast with the transition between the plateaus, here $V_T$ switches between positive and negative sign over a small range of $V_G$ (Figure~3a). This behavior is reminiscent of the mesoscopic fluctuations of the downstream resistance due to CAES found in Ref.~\cite{zhao_interference_2020}, which has been attributed to the effect of the random distribution of vortices in the superconducting contact. Similar to that case, we find that the pattern of $V_T$ strongly depends of the history of magnetic field: while patterns are very stable as $V_G$ is swept at fixed $B$, sweeping the field at fixed $V_G$ results in very random traces, Figure~3c. Here, the signal switches sign every few mT, which can likely be attributed to the rearrangement of the vortices. After sweeping the magnetic field, the original pattern of $R_T (V_G)$ is not recovered.
\\
\begin{figure}[h]
\centering
\includegraphics[width=\textwidth]{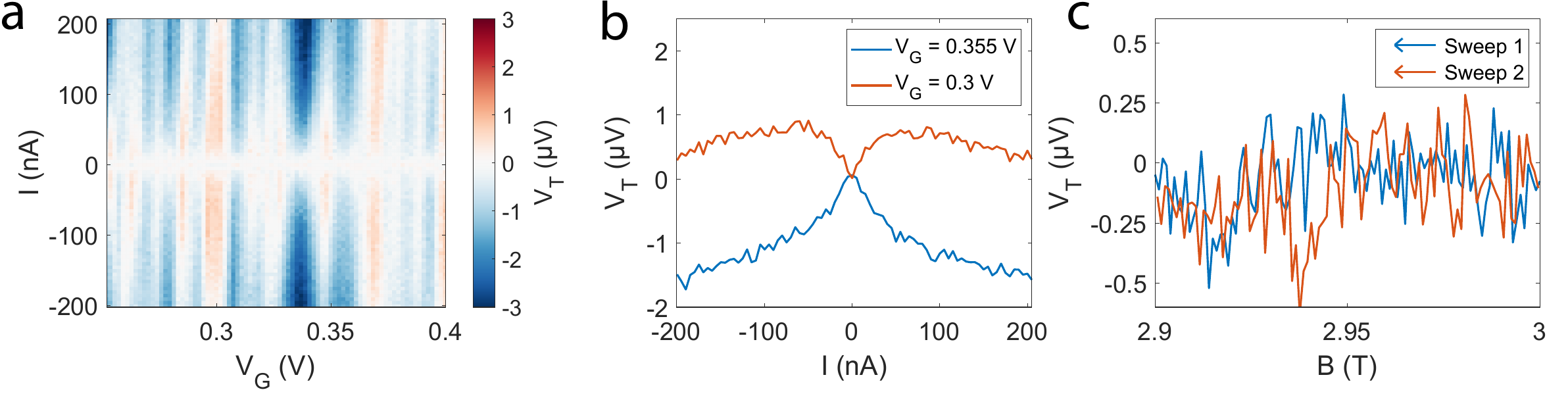}
\label{fig:vortex}
\caption{\textbf{(a)} $V_T$ measured as a function of $I$ and $V_G$ on the $\nu=4$ plateau at 2 T in the 250~nm device. \textbf{(b)} Vertical cuts of \textbf{a} reveal a prominent `V' shaped curve of both signs, indicative of thermopower effects. \textbf{(c)} Traces of $V_T$ taken at $I=100$ nA in the $W=50$ nm device. The field is first swept from $B=3$ to 2.9 T and then further to $B=2.7$ T. The $B=3\rightarrow 2.9$ T trace is then repeated. The previously repeatable measured $V_T$ is irrecoverable, indicative of the influence of vortices in the thermal transport.}
\end{figure}

\begin{figure}[h]
\centering
\includegraphics[width=\textwidth]{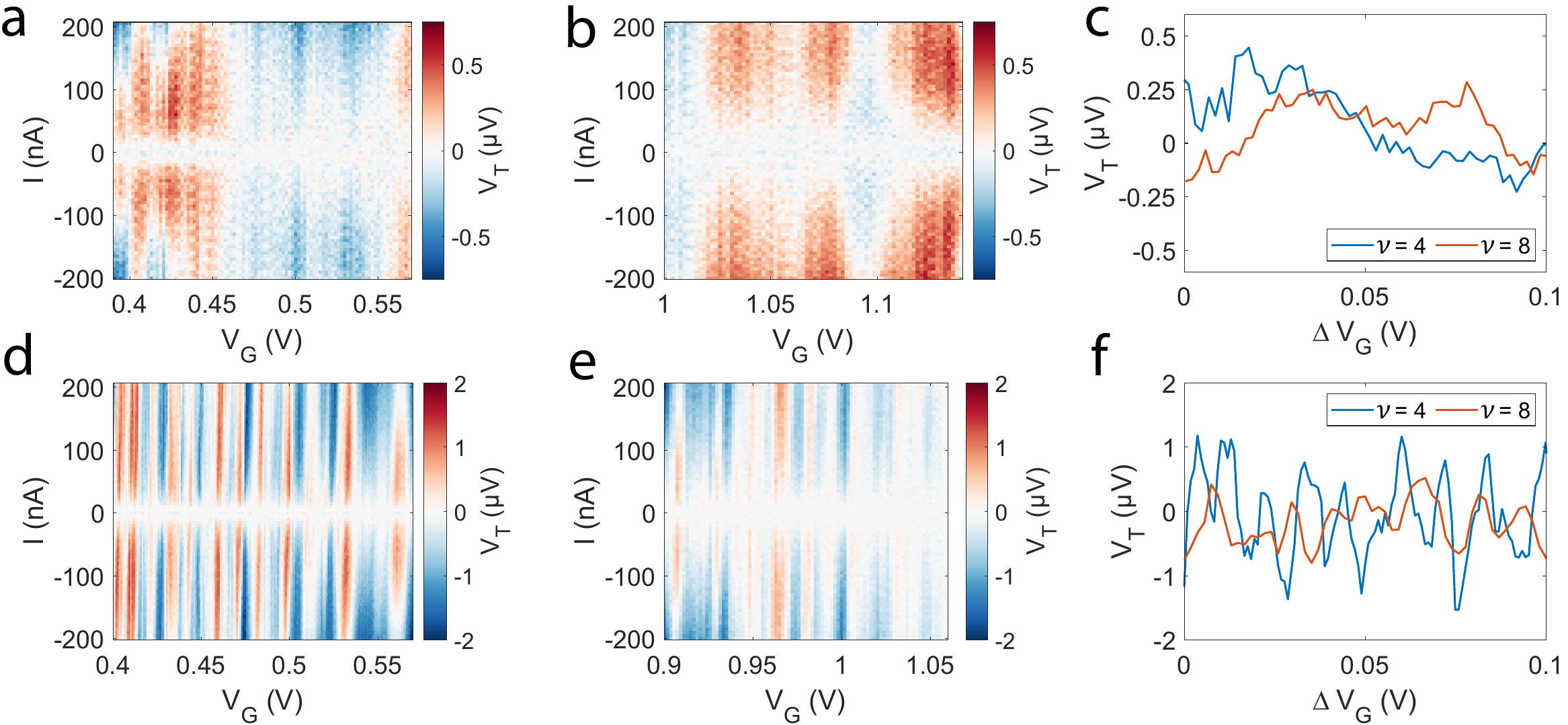}
\label{fig:ThickAndThin}
\caption{\textbf{(a-b)} $V_T$ measured vs bias and gate voltage in the 50~nm device across the $\nu=4$ and $\nu=8$ plateaus respectively. \textbf{(c)} Trace of \textbf{a-b} taken at 100 nA and beginning at the left of each plot. \textbf{(d-f)} Corresponding measurements in the 250~nm device.}
\end{figure}

\noindent Finally, in Figure~4 we compare two filling factor, $\nu=4$ and 8, and two devices. In a given device (top and bottom rows), the signals are qualitatively similar for the two filling factors. However, the oscillations of $V_T$ vs. $V_G$ are more frequent in the wider wire (250~nm, panels d-f). The signal is also consistently larger in the wider wire. The enhanced amplitude of $V_T$ is somewhat counterintuitive given that one may expect the narrower wire (50~nm) to be hotter. We discuss possible reasons for this behavior in the next section. Or not.

\section{Discussion}
We first address the $V_T(I)$ curves which appear to have a `V' shape at small bias (Figure~3b). We equate the Joule heating power $P_{in} = \frac{I^2h}{\nu e^2}$ with the cooling power. We have measured the latter to follow the same functional dependence as the normal electrons, $P_{out} = \alpha(T^2-T^2_0)$, where $T$ is the hot electron temperature and $T_0$ is the base temperature  \cite{zhao_thermal_2025}. For sufficiently small $T_0$, equating $P_{in}=P_{out}$ results in
$T\propto |I|$. Finally, the hot electron distribution is expected to propagate to the opposite side of the wire, resulting in the thermopower $V\propto |I|$. This linear dependence is expected to saturate when the electron temperature $T$ reaches a few hundred mK, at which point the electron cooling via emission of phonons becomes efficient. 

We now focus on the measurements conducted on the plateau. 
The switching in sign of $V_T$ implies the preferential emission of electrons or holes from the contact. This suggests a complicated density of subgap states, likely originating from vortex cores~\cite{caroli_bound_1964}. Sweeping the gate voltage changes the way the edge states equilibrate with the hot distribution in the cores, resulting in variations of $V_T$. The highly stochastic dependence of the signal on magnetic field further supports the role of vortices in this phenomenon. \\
\\

\noindent Finally, we note that our results are not consistent with Andreev reflection across the superconducting wire, as proposed in~\cite{clarke_exotic_2014}. The arguments against such interpretation include the clear role played by the vortices, the even shape of the $V_T(I)$ curves, and the fact that the signal is enhanced in the wire with $W=250$~nm, which greatly exceeds the coherence length. Furthermore, the graphene region is fully cut by the central contact, precluding any edge state from propagating around the superconductor. These arguments collectively indicate that the observed phenomena are explained by
thermal physics, rather than quantum Hall superconducting proximity. 

\section{Conclusion}
In summary, we have measured the voltage across a superconducting wire produced by quantum Hall edge states incident on a superconducting interface. The resulting signal displays a clear `V-shaped' non-local $I\textendash V$ curves which we attribute to a thermopower effect produced by a hot electron distribution. The signal displays a clear dependence on the presence of chiral edge states and vortex configuration.  Alternative explanations based on Andreev reflections across the superconductor appear inconsistent with these observations.  Thermal signals may complicate the search of exotic excitations at the interfaces of quantum Hall and superconductors.\\
\\

\backmatter

\bmhead{Author Contributions}

The device was designed by J.T.M and G.F. The fabrication and measurement of the device was performed by J.T.M. The deposition of MoRe was performed by F.A. K.W and T.T provided the hBN crystals.

\bmhead{Funding}
The work at Duke University was supported by the Division of Materials Sciences and Engineering, Office of Basic Energy Sciences, U.S. Department of Energy, under Award No. DE- SC0002765. Sample fabrication was performed in part at the Duke University Shared Materials Instrumentation Facility (SMIF), a member of the North Carolina Research Triangle Nanotechnology Network (RTNN), which is supported by the National Science Foundation (award number ECCS-2025064) as part of the National Nanotechnology Coordinated Infrastructure (NNCI). F.A. acknowledge support from ARO (award W911NF-16-1-0132). K.W. and T.T. acknowledge support from JSPS KAKENHI grant no. JP15K21722 and the Elemental Strategy Initiative conducted by the MEXT, Japan. T.T. acknowledges support from JSPS Grant-in-Aid for Scientific Research A (no. 26248061) and JSPS Innovative Areas Nano Informatics (no. 25106006).
\bmhead{Acknowledgements}
We thank Ethan Arnault for useful discussions whilst preparing the manuscript.

\bigskip

\begin{appendices}

\section{Device Fabrication}

The heterostructure was assembled from flakes of graphene, graphite and hBN exfoliated on a diced silicon wafer with a thermally grown 280~nm oxide layer. Flakes of graphene, graphite and hBN of suitable thickness are identified optically prior to heterostructure assembly. The flakes stacked in to  a heterostructure with the dry transfer technique and picked up with a polydimethylsiloxane (PDMS)/polyethylene terephthalate (PET) stamp. The PDMS hemisphere was defined by drop casting on to a hot glass slide, next a 2x2cm PET square was placed over the PDMS and baked at 180c for 10 mins to promote adhesion.\\
\\
Individual flakes were picked up with the stamp at 60°C and then deposited onto a clean Si/SiO$_2$ substrate at 180°C. The finished stack was cleaned of stamp residue in hot dichloromethane for 10 minutes, followed by an overnight anneal step at 500°C in atmosphere. A clean region of encapsulated graphene was identified and tip cleaned via atomic force microscopy.\\
\\
Patterning of the heterostructure was achieved by electron-beam lithography on a layer of PMMA resist. A calibrated CHF3/O2 etch was used to make one-dimensional contact to the graphene layer, without electrically shorting to the graphite back gate.  The superconducting electrodes comprised 100nm of MoRe alloy (50–50 ratio by weight) d.c. sputtered in a high-vacuum chamber (10$^{-8}$ torr). Contact to the graphite back gate was made by etching through the whole stack and then thermally evaporating Cr/Au (1/90nm) in a vacuum of $\approx10^{-7}$ torr.

\section{Measurement}
The transport measurements were carried out in a Leiden Cryogenics dilution refrigerator. The measurement lines were fitted with 2nd-order low-pass filters, thermalized to the mixing chamber plate, to avoid spurious microwave frequency radiation from heating the sample. Measurements were performed in a quasi-four-probe configuration.

\end{appendices}

\bibliography{QHSCThinWireHeatingRefs}

\end{document}